\documentclass[12pt,preprint]{aastex}
\usepackage{graphicx}
\usepackage{wasysym} % for a large circle symbol
\newcommand\ba{\begin{eqnarray}}
\newcommand\ea{\end{eqnarray}}
\newcommand\be{\begin{equation}}
\newcommand\ee{\end{equation}}

\newcommand\rms{\mathrm{s}}
\newcommand\rme{\mathrm{e}}
\newcommand\rmi{\mathrm{i}}
\newcommand\rmp{\mathrm{p}}

\newcommand\rmK{\mathrm{K}}

\newcommand\rmg{\mathrm{g}}

\newcommand\AU{\mathrm{AU}}
\newcommand\gm{\mathrm{gm}}
\newcommand\cm{\mathrm{cm}}
\newcommand\mm{\mathrm{mm}}
\newcommand\mfp{\mathrm{mfp}}

\newcommand\p{\partial}
\newcommand\ts{t_\rms}

\newcommand\uh{\widehat u}
\newcommand\vh{\widehat v}
\newcommand\Dh{\widehat D}
\newcommand\sigmah{\widehat \sigma}
\newcommand\sansA{\mathsf{A}}
\newcommand\taus{\tau_\rms}

\newcommand\QT{Q_\mathrm{T}}
\newcommand\QR{Q_\mathrm{R}}

\newcommand\calT{\mathcal{T}}

\newcommand\uk{u_\rmK}
\newcommand\rhog{\rho_\rmg}

\def\drawline#1#2{\raise 2.5pt\vbox{\hrule width #1pt height #2pt}}
\def\spacce#1{\hskip #1pt}
\newcommand\solid{\drawline{24}{.5}\nobreak\ }
\newcommand\bdash{\hbox{\drawline{4}{.5}\spacce{2}}}

\newcommand\dashed{\bdash\bdash\bdash\bdash\nobreak\ }

\newcommand\bdot{\hbox{\drawline{1}{.5}\spacce{2}}}
\newcommand\dotted{\hbox{\leaders\bdot\hskip 24pt}\nobreak\ }
\newcommand\chndot{\hbox {\drawline{9.5}{.5}\spacce{2}\drawline{1}{.5}\spacce{2}\drawline{9.5}{.5}}\nobreak\ }
\newcommand\chnddot{\hbox {\drawline{9.25}{.5}\spacce{2}
                    \drawline{1}{.5}\spacce{2}
                    \drawline{1}{.5}\spacce{2}
                    \drawline{9.25}{.5}}\nobreak\ }

\def\eqp#1{(\ref{eq:#1})}

\shorttitle{Gravitational instability of solids}
\shortauthors{Shariff \& Cuzzi}

\begin{document}


\title{Gravitational instability of solids assisted by gas drag: slowing by
turbulent mass diffusivity} 


\author{Karim Shariff and Jeffrey N. Cuzzi}
\affil{NASA Ames Research Center, Moffett Field, CA 94035 \\
{\rm (Accepted, June 14, 2011, Astrophysical Journal)}}

\begin{abstract}
The \cite{GW73} (axisymmetric) gravitational instability of a razor
thin particle 
layer occurs when the Toomre parameter $\QT \equiv c_\rmp \Omega_0 / \pi G \Sigma_\rmp < 1$
($c_\rmp$ being the particle dispersion velocity).
\cite{Ward76,Ward00} extended this analysis by adding the effect of gas drag upon particles and
found that even when $\QT > 1$, sufficiently long waves were always unstable.
\cite{Youdin05a, Youdin05b} carried out a detailed analysis and
showed that the instability allows chondrule-sized
($\sim 1 $ mm) particles to undergo radial clumping with reasonable
growth times even in the presence of a moderate amount of turbulent
stirring.  The analysis of Youdin includes the role of turbulence
in setting the thickness of the dust layer and in creating a turbulent
particle pressure in the momentum equation.
However, he ignores the
effect of turbulent mass diffusivity on the disturbance wave.  Here we show 
that including this effect reduces the growth-rate significantly, by an amount 
that depends on the level of turbulence, and reduces the maximum intensity of
turbulence the instability can withstand by 1 to 3 orders of magnitude.  The
instability is viable only when turbulence is extremely weak
and the solid to gas surface density of the particle layer is considerably enhanced over
minimum-mass-nebula values.  A simple mechanistic
explanation of the instability shows how the azimuthal component
of drag promotes instability while the radial component hinders it.
A gravito-diffusive overstability is also possible but never realized in the
nebula models.
\end{abstract}



\section{Introduction}

\subsection{Preliminary Remarks}

It is believed that clumping of solid material to form the terrestrial planets and the putative cores of gas giants involved three stages.  The first and third stages are relatively well understood.  First, grains that survived shocked entry into the solar nebula and those that condensed from the cooling gas, collided and stuck by van der Waals attraction; such a process is considerably speeded up by
turbulence but is still effective in laminar disks \citep{Weid80,Weid84,DD05} and is able to form cm-or-larger sized particles. Thereafter, growth by binary accretion becomes problematic in a turbulent nebula due to shattering. 
In the third stage, planetesimals, bodies 1 km and larger which are akin to the present day asteroids, merged in binary fashion by physical and/or gravitational capture as they collided \citep{KI00,Cham01,KenBrom01}.

Least understood is the middle stage, namely, the growth from cm to
km-sized bodies.   Early on \cite{Safronov72} and
\cite{GW73}(henceforth collectively referred to as SGW) independently
suggested that particles settled to the midplane and underwent a
gravitational instability (GI).  This caused clumps of some characteristic size to contract until centrifugal force became strong enough to balance self-gravity.  \cite{GW73} suggested that bodies of size $\lesssim 0.5$ km having the density of solid material could form in this way on a dynamical timescale.

The SGW scenario ignored the presence of global turbulence in the
solar nebula which stabilizes the dust layer against gravitational
instability by making the basic state particle layer more diffuse and introducing
a turbulent pressure into the particle dynamics.
A possible source of disk
turbulence is magneto-rotational instability (MRI).  It has been
argued \citep{Gammie96} that MRI may be confined to only the upper
layers of a disk which are sufficiently ionized by cosmic
rays. Nevertheless, this does not imply a completely quiescent
mid-plane since vortical eddies in the unstable layers can induce
fluctuations at the midplane.
\cite{DD04} found that laminar settling of dust grains could not account for the observed flaring of disks suggesting that turbulence is keeping small particles aloft throughout the disk height.

Even if one discounts global turbulence, an obstacle to the SGW picture of
midplane GI remains: the dust disk itself can generate turbulence.  The gas
component feels a small outward directed pressure force and therefore rotates at
a slightly slower speed than the Keplerian speed of solid particles.  This
creates an Ekman-like layer whose density stratification is not strong enough to
stabilize the layer.  The turbulent state is such that the critical particle
density required for the mid-plane gravitational instability is not realized
\citep{Weid80,Weid84, Cuzzi_etal93}.

A variant of midplane gravitational instability involves suppression of turbulence in
the mixed particle-gas layer by the vertical density gradient. This variant
is valid only for very small particles, which are so well
trapped to the gas that the particle-gas mixture behaves as a single fluid. In
this limit, \cite{Sekiya98} calculated the (mean) vertical structure of the
turbulent dust layer by assuming that the mean profiles of density and velocity
lie at the instability/stability boundary, {\it i.e.,} characterized by a
Richardson number equal to its critical value of $1/4$.  He found that if the
ratio of solid to gas surface density in the disk is much larger than the cosmic
value, the layer achieves the critical density needed for GI.  This type of
analysis produces a cusp at the midplane in the mean density profile as the
surface density of solids in the disk is increased.
\cite{Youdin_and_Shu02} and later \cite{YC04} studied this situation further, 
suggesting that the
cusp was not a mathematical artifact but indicative of the inability of the
turbulence to stir up particles near the midplane.  They proposed that the
required enrichment of solids could occur as particles drift inward by gas
drag. \cite{Garaud_and_Lin04} studied the Richardson number
instability as the particle layer slowly settles.  They also find (their Fig.
11) that, at the levels of the pressure force appropriate to a minimum mass
nebula, gravitational instability sets in before the Richardson number
instability only if the column density of particles is $\approx 1/5$ the column
density of the gas; this represents a factor of $\approx 30$ increase over the
minimum-mass-nebula model, in agreement with previous works. Because of
the small particle limitation, this variant of midplane GI is precluded by
even the faintest breath of global turbulence \citep{CW06}.  

Alternate scenarios of growth in dense
midplane layers, which do not rely on gravitational instabilities, have been
developed by Weidenschilling in a number of papers \citep[see][for a
review]{Weid00}.  A review of this stage is given by \cite{CW06}
and recent work \citep{Cet01,JKH06}  suggests that the problem is more complex than 
envisioned in the early works.

\subsection{The gas-drag mediated instability}

The subject of the present work is a gravitationally driven instability made possible by gas drag.  What makes the instability interesting is that, although it is relatively slow, it is unconditional.
The instability was proposed and studied by
\citet{Ward76,Ward00} in the absence of turbulence and associated particle
dispersion. \cite{Safronov87,Saf91} also briefly considers this instability
under the same conditions and writes down the dispersion relation in the limit
where particle stopping time is short compared to the time scale of the
perturbation.  A detailed analysis \citep[][henceforth papers I and
II]{Youdin05a, Youdin05b} suggested that the gas-drag assisted instability remains
viable in the presence of a moderate amount of nebula turbulence.  This analysis
included the effect of turbulence in the particle momentum equation (via an
effective pressure) and in setting the height of the disk layer.  However, a
turbulent diffusion term must also be present in the equation for particle mass
conservation. Here we correct this omission and find significant reductions in
growth-rate.  In particular, we find that the instability is not viable for
nominal values of nebula turbulence unless favorable assumptions are made
regarding local conditions.  These include some combination of elevated solid/gas ratio, and
particle size neither too small nor too large.

During review of the revised version of this paper, we were sent a new paper, \cite{Youdin11},
that also corrects for the lack of turbulent mass diffusivity in the 2005 papers.  In addition,
it uses updated models for the radial turbulent diffusivity of particles, the radial particle dispersion,
and the height of the particle sub-disk. 
These models are based on the analysis of 
\cite{Youdin_and_Lithwick07} 
and the corrections they
imply become important for particles whose stopping times $\ts$ non-dimensionalized by the
orbital frequency $\Omega_0$ are such that $\taus \equiv \Omega \ts \gtrsim1$. 
The present calculations were therefore redone to incorporate the newer models, however, the
basic conclusions remained the same.

\section{Analysis}

\subsection{Dispersion Relation}

When a mass diffusion term with coefficient $D$ is included, the linearized and vertically
integrated mass and 
momentum equations for the particles, equations (6) and (7) in paper I, become 
\ba
  \frac{\p\sigma}{\p t} + \frac{\p u}{\p x} &=& D \frac{\p^2\sigma}{\p x^2}, \label{eq:cons1}\\
  \frac{\p u}{\p t} - 2 v \Omega_0 &=& - \frac{\p\chi}{\p x} - \frac{u}{\ts}, \label{eq:cons2} \\
  \frac{\p v}{\p t} + (2 - q) u \Omega_0 &=& - \frac{v}{\ts}, \label{eq:cons3}
\ea
where $q \equiv 3/2$.
The equations are written relative to a box at radius $R$ revolving at the local angular
velocity $R\Omega(R)$ of particles; we have defined $\Omega_0 \equiv \Omega(R)$.  
Affixed to the box are local Cartesian coordinates 
$(x, y)$, where $x$ is radial and $y$ is azimuthal. 
The corresponding velocity components are $(u, v)$: they represent turbulent mean
quantities that have then been vertically averaged.
The quantity $\sigma$ is the relative perturbation in particle surface density
defined so that:
\be
   \Sigma = \Sigma_\rmp \left(1 + \sigma\right),
\ee
where $\Sigma_\rmp$ and $\Sigma$ are the basic state and total (basic state $+$ perturbation) 
surface densities, respectively.  Throughout the analysis used to 
obtain \eqp{cons1}--\eqp{cons3},
one assumes that $\Sigma_\rmp$ is locally uniform and can therefore be freely moved into and
out of $x$ and $y$ derivatives.  
The basic state velocity of particles has been 
taken to be Keplerian in deriving \eqp{cons1}--\eqp{cons3}.
The gas flow has also been taken to be Keplerian and is left unperturbed.  In this model
therefore, there is only one-way coupling from the gas to the particles.  (One consequence
of two-way coupling are streaming instabilities, e.g., \cite{Youdin_and_Goodman05}).
Figure 3 (lower) in \cite{Cuzzi_etal93} shows that the turbulent mean flow departs from being
Keplerian
by only about 0.2\%.  The basic state flow of the particles should also have an inward radial
drift arising from gas drag \citep{Adachi_etal76}; this has been left out 
in \eqp{cons1}--\eqp{cons3}.  
However, since the (vertically averaged) drift velocity is locally uniform with respect 
to $x$ and $y$, it would cause the
instability wave to merely drift inward without affecting its local growth-rate.
Effects that involve the fact that the drift velocity has a vertical dependence are not
included in the analysis.
We found (except for the gravito-diffusive mode that was never realized in the nebula calculations)
that the phase and group velocity of the instability wave is zero in the
drifting frame.  This comes about simply because the wave oscillation frequency is zero
in some neighborhood of the most amplified wavenumber.  Thus,
the wave drift speed is the same as the particle drift speed and this will limit the
total amount of wave growth.
This is accounted for in the study
by comparing the e-folding growth time to a local drift time scale
\be
   t_\mathrm{drift} = R / u_\mathrm{drift}.
\ee
Complete loss of solids from the entire nebula is a weaker constraint.

Axisymmetric disturbances are being considered.  Thus, the
only spatial derivatives that appear are those with respect to the radial
coordinate $x$.

The quantity $\chi \equiv \Phi + \Pi'$ consists
of the gravitational potential $\Phi$ and $\Pi' \equiv \sigma c_\rmp^2$ which
arises from modeling the effect of turbulence on particle momentum by an effective pressure, $c_\rmp$
being the radial particle dispersion velocity 
\citep{Youdin_and_Lithwick07}.  
The effective pressure term arises from the
$rr$ component of the particle Reynolds stress tensor in the Reynolds-averaged particle momentum equation.
It should be noted that neither the present treatment nor that of \cite{Youdin11} has incorporated all
the effects of turbulence in the particle momentum equation.  A Reynolds average of the original
equation reveals that, in addition to an effective pressure, several other turbulent correlations arise that 
require closure models.
Compared to previous treatments 
\citep{Ward00, Youdin05a, Youdin05b} the only new term is the 
right hand side of \eqp{cons1} and represents turbulent diffusion of
surface density.  
This term arises from the correlation $\left<\rho^\prime u_j^\prime\right>$ in the Reynolds averaged particle mass conservation
equation (the primes denote turbulent fluctuations and the angle
brackets denote a suitable Reynolds average). 
This has been modeled using gradient diffusion as $-D\partial \rho/\partial x_j$.

Substituting into equations \eqp{cons1}-\eqp{cons3} perturbations of the form 
\be
(\sigma, u, v) = (\sigmah, \uh, \vh) \rme^{\rmi(kx - \omega t)},
\ee
where $(\sigmah, \uh, \vh)$ are complex constants, gives a linear homogeneous 
system
\be
\sansA (\sigmah, \uh, \vh)^{T} = 0,
\ee
where $\sansA$ is a matrix.
The solvability condition for this system, namely, that the determinant of $\sansA$
vanish, gives the dispersion relation:  
\be
   \left( \omega + \rmi/\ts\right)
   \left[ C(k) -  (\omega + \rmi D k^2)(\omega + \rmi / \ts)\right] +
   \rmi\Omega_0^2 (Dk^2 - \ts^{-1}) = 0,  \label{eq:disp}
\ee
where
\be
   C(k) \equiv \Omega_0^2 + k^2 c_\rmp^2 - 2 \pi G \Sigma_\rmp k \calT(kh_\rmp). \label{eq:B}
\ee
Here $h_\rmp$ is the height of the particle sub-disk and
$\calT(k h_\rmp) = 1/ (1 + k h_\rmp)$ is a factor that 
approximates finite thickness effects on the potential of self-gravity
in the context of a vertically integrated model.
In the limit $\ts \to \infty$ (vanishing drag) and $D \to 0$,
\eqp{disp} reduces to the Safronov-Goldreich-Ward (SGW) form $\omega^2 = C(k)$.
Since turbulent mass diffusivity occurs only in the product $Dk^2$, its effect is arbitrarily small for sufficiently long waves.  However, the most amplified wave has a finite $k$ and therefore 
turbulent mass diffusivity has a non-negligible effect on growth-rate.

Following \cite{Youdin05a} let us introduce the non-dimensional variables
\be
   \gamma \equiv - \rmi \omega / \Omega_0, \hskip0.5truecm
   \kappa \equiv \pi G \Sigma_\rmp k / \Omega_0^2, \hskip0.5truecm
   \taus \equiv \Omega_0 \ts, \hskip 0.5truecm
\ee
and the Toomre and Roche parameters
\be
   \QT \equiv \frac{c_\rmp \Omega_0}{\pi G \Sigma_\rmp}, \hskip0.5truecm
   \QR \equiv \frac{h_\rmp \Omega_0^2}{\pi G \Sigma_\rmp}.
\ee
The additional parameter resulting in the present case is the non-dimensional
diffusivity:
\be
   \Dh \equiv \frac {D \Omega_0^3}{(\pi G \Sigma_\rmp)^2}.
\ee
The quantity $\mathrm{Re}(\gamma)$ gives the growth-rate and
the parameter
$\QR$ is present only because finite thickness effects on self-gravity have
been retained.  The dispersion
relation \eqp{disp} may then be written as:
\be
\left( \gamma + \tau_\rms^{-1} \right)
\left[  F + \left(\gamma + \Dh \kappa^2\right)\left(\gamma + \tau_\rms^{-1}\right) \right]
+ \Dh \kappa^2 - \tau_\rms^{-1} = 0, \label{eq:disp_nd}
\ee
where
\be
   F(\kappa, \QT, \QR) \equiv 1 - \frac{2 \kappa}{1 + \kappa \QR} + \QT^2 \kappa^2. 
\ee
As can be checked, the condition $F < 0$ gives instability in the limit of zero gas drag ($\taus \to \infty$) and zero mass 
diffusivity ($\Dh \to 0$).
Equation \eqp{disp_nd} expands to the following cubic equation for the $\gamma$:
\be
   \gamma^3 + \left(2\taus^{-1}+\Dh\kappa^2\right)\gamma^2 + \left(\taus^{-2}+F+2\Dh\kappa^2\taus^{-1}\right)\gamma
   + \taus^{-1}\left(F-1\right) + \Dh\kappa^2\left(1+\taus^{-2}\right) = 0.
\ee
Using Descartes' rule of signs and clever reasoning, \cite{Youdin05a} concluded that a necessary and sufficient condition for 
instability (for the case of zero mass diffusivity) is that $F(\kappa, \QT, \QR) - 1 <
0$.  A similar analysis does not appear to be possible in the present case.  However, \cite{Youdin11} has shown that,
even with mass diffusivity,
one can always find a wavenumber $\kappa$ such that the system is unstable.

The slow instabilities considered are such that the growth-rates can be much
smaller than the particle stopping rate, i.e., $\gamma \ll 1/\taus$.  In this limit,
an explicit expression is obtained for the growth-rate from \eqp{disp_nd}:
\be
   \gamma_\mathrm{approx} = \taus(1 - F)  - \Dh \kappa^2 \left(\taus^2 + 1\right),
   \label{eq:approx}
\ee
which clearly shows the damping effect of $\Dh$.  The validity of
\eqp{approx} can be checked \textit{a posteriori}.

\subsection{Implementation}

To implement the analysis for various disk conditions we follow exactly the treatment of
papers I and II updated with the models in \cite{Youdin11} for the effect of turbulence on
particles.  Defining $\varpi \equiv R/\AU$, the nebula model employed is
\begin{eqnarray}
   \Sigma_\rmg &=& 1700 f_\rmg \varpi^{-3/2}\ \gm\ \cm^{-2},\\
   \Sigma_\rmp &=& 10     f_\rmp \varpi^{-3/2}\ \gm\ \cm^{-2}, \\
   c_\rmg &=& 10^5 \varpi^{-1/4}\,\cm\ \rms^{-1},
\end{eqnarray}
for the gas-disk surface density, particle-layer surface density, and sound speed, respectively.
Apart from the factors $f_\rmg$ and $f_\rmp$, this is just the minimum mass
model.
The calculation of the non-dimensional stopping time, $\taus$, uses the Epstein and Stokes
formulas as appropriate:
\be
\taus = \left\{\begin{array}{ll}
4\times 10^{-4} (a/\mm) \varpi^{3/2} f_\rmg^{-1}, & a/\lambda_\mfp \le 2.002;\\
9\times 10^{-5} (a/\mm)^2 (\varpi / 0.3)^{-5/4}, & \mathrm{otherwise},\\
\end{array}\right. \label{eq:drag}
\ee
where
\be
   \lambda_\mfp = f_\rmg^{-1} \varpi^{2.75},
\ee
is the mean free path.
The resulting $\taus$ depends on both particle size and radial location
in the disk.  Equation (\ref{eq:drag}) is valid when the particle Reynolds number $\mathrm{Re}_D = 2a \Delta u/\nu \lesssim 1$.
This condition is satisfied in all the plots we present by adjusting the range 
of the abscissa if necessary.
Here $\Delta u$ is the magnitude of the particle velocity relative to the gas which depends on $\taus$ according to equation
(A1) in \cite{Youdin11}, and $\nu = 2.45\times 10^4 \varpi^{5/2} f_\rmg^{-1}$ cm$^2$ s$^{-1}$ is the kinematic viscosity. 

The radial component of the particle dispersion velocity, $c_\rmp$, is calculated using
equation (27) in \cite{Youdin11}:
\be
c_\rmp = \frac{(1 + 2\taus^2 + (5/4)\taus^3)^{1/2}}{1 + \taus^2}\sqrt{\alpha_\rmg} c_\rmg.
\ee
The radial mass diffusivity due to turbulence is written as:
\be
  D = \frac{\nu_\rmg}{\mathrm{Sc_T}},
\ee
where $\nu_\rmg$ is the turbulent momentum diffusivity of the gas and $\mathrm{Sc_T}$
is the turbulent Schmidt number.  Following \cite{Youdin11}
\be
  \mathrm{Sc_T} = \frac{(1 + \taus^2)^2}{1 + \taus + 4\taus^2}.
\ee  
The turbulent viscosity of the gas, $\nu_\rmg$, is defined via an
$\alpha$ parameter:
\be
 \nu_\rmg = \alpha_\rmg c_\rmg^2 / \Omega_0,
\ee
where the turbulence parameter $\alpha_\rmg$ is a measure of the local turbulence
intensity of the gas and $c_\rmg$ is the sound speed of the gas.
After some substitutions one obtains:
\be
   \Dh = \alpha_\rmg
   \frac{1 + \taus + 4\taus^2}{(1 + \taus^2)^2}
   \left(\frac{Q_\mathrm{Tg} \Sigma_\rmg}{\Sigma_\rmp}\right)^2,
   \label{eq:Dh} 
\ee
where $Q_\mathrm{Tg}$ is the Toomre parameter of the gas disk:
\be
   Q_\mathrm{Tg} \equiv \frac{c_\rmg \Omega_0}{\pi G \Sigma_\rmg}.
\ee
For the present nebula model we obtain:
\be
\left(\frac{Q_\mathrm{Tg} \Sigma_\rmg}{\Sigma_\rmp}\right)^2 = 
10^8 f_\rmp^{-2} (\varpi)^{-1/2},
\ee
whose large leading coefficient accounts for the sensitivity to $\alpha_\rmg$ observed in the results.

The model of \cite{Youdin11} used to determine the height $h_\rmp$ of the particle disk is:
\be
   h_\rmp = h_\rmg \sqrt{\frac{\alpha_\rmg}{\taus\psi + \alpha_\rmg}}, \label{eq:hp}
\ee
where
\be
   \psi \equiv 1 + \frac{2\pi G \Sigma_\rmp}{\Omega_0^2 h_\rmp} = 1 + 2/\QR. \label{eq:psi}
\ee
Equations \eqp{hp} and \eqp{psi} lead to a quadratic equation for $h_\rmp/h_\rmg$ which is
explicitly solved.

Finally, we discuss the constraint due to inward radial drift of the
instability wave as motivated earlier.  We have
\be
   t_\mathrm{drift} = R/ |u_\mathrm{drift}|.
\ee
The analysis of \cite{Nakagawa_etal86}, extended to include terms quadratic 
in $\taus$, gives for the equilibrium
(i.e., neglecting acceleration terms) radial drift speed of the particle layer:
\be
   u_\mathrm{drift} = -2\eta u_\rmK \left[
   \frac{\taus}{(1 + \phi_\rmp(z))^2 + \taus^2}
   \right],
\ee
where $\uk = \Omega_0 R$ is the Keplerian speed,
\be
   \eta \equiv - \frac{R}{2\rhog\uk^2}\frac{\partial p}{\partial R} = 1.3\times 10^{-3} \varpi^{1/2},
\ee
is the pressure gradient parameter, and 
$\phi_\rmp(z) \equiv \rho(z)/\rhog(z)$ is the ratio of particle to gas density (the so-called particle loading) which in general depends on the vertical position $z$ in the disk.
Using vertical averages for both numerator and denominator we estimate
\be
   \phi_\rmp \equiv \frac{\rho(z)}{\rhog(z)} \approx \frac{\Sigma}{\Sigma_\rmg}\frac{h_\rmg}{h}. 
\ee

\subsection{Mechanistic Interpretation of the Instability}
\label{sec:mech}

Before we begin let us note that gas drag (identified by the parameter $\taus$)
enters not only as an explicit term in the momentum equations but also through 
setting the turbulent diffusivity, effective pressure, and height of the particle layer.  
That is, $\taus$ also affects the parameters $\Dh$, $\QT$, and $\QR$.
In this sub-section we focus only on the role of the explicit gas drag terms in the 
momentum equations.

Gravitational instability occurs when the increase of mutual gravity
between approaching particles outweighs restoring forces.
As is well known, an important restoring force in Keplerian systems and for axisymmetric
perturbations comes from the outwardly increasing angular momentum.
Consider a perturbation mode (in radial velocity and density) such that particles 
accumulate in a toroidal region and deplete in the neighboring region.  
Self-gravity is not needed at this stage.
A circle of particles at the inner edge of this region is displaced outward.
If gas drag is absent, the circle will conserve its specific
angular momentum $r u_\theta$ and it will slow down more steeply than the local
Keplerian speed slows with radius.  It will therefore have a deficit of centripetal
acceleration relative to the pull of the central gravitator and will experience an
inward restoring force.
Similarly, a circle at the outer edge of the toroidal region will be displaced 
inward and will speed up
faster than the local Keplerian value and experience an outward
restoring force.  This restoring force is what makes Keplerian disks
stable via Rayleigh's criterion and is characterized by a $+\Omega_0^2$ term
on the right-hand-side of the dispersion relation $\omega^2 = C(\omega)$ in the drag-free (SGW)
case.

Let us now consider the same picture but in the presence of the azimuthal gas drag term
in the momentum equation.
In this case, the azimuthal speed of an outward/inward displacing particle circle will
remain closer to the local Keplerian value and there will not be as much of
a deficit/excess of centripetal acceleration compared to the pull 
of the central gravitator.  
The restoring force is thus diminished.  
This is the mechanism by which the azimuthal gas drag term
promotes instability.  Ultimately, what has to overcome whatever restoring force 
remains is the increase in mutual gravity among the accumulating particles.

Note that for an outwardly displacing circle, the gas flow is
a tailwind relative to the particles and gas drag supplies them with 
\textit{more angular momentum}.

In actual fact, the gas revolves slightly slower than Keplerian due to pressure support.  
Hence an outwardly/inwardly displaced circle will actually experience a slightly 
smaller tailwind/larger headwind than envisioned above.  The net result is that 
everything that was described in the preceding paragraphs 
should be imagined to be taking place in a reference frame drifting inward
(in the vertically averaged model).

Consider the limit of large drag; specifically, assume that the stopping time $t_\rms$ 
is very short compared to the time scale $1/|\omega|$ of the perturbation.  In this
case the azimuthal speed of the displaced circle will always remain very
close to that of the local gas, namely, Keplerian.  In  this limit, therefore,
the deficit/excess of angular momentum of displaced circles and hence
the restoring tendency is completely anulled.  One can see this explicitly
by writing the dispersion relation \eqp{disp} for this case:
\be
   \ts^{-1}\left[C(k) - (\omega + \rmi D k^2)\,\rmi\,\ts^{-1}\right] +
   \Omega_0^2 D k^2 - \Omega_0^2 \,\ts^{-1} = 0. \label{eq:cancellation}
\ee 
One observes that the restoring force term ($\Omega_0^2$
in $C(k)$) is cancelled by the last term in \eqp{cancellation}.

So far, our discussion has only considered
the role of gas drag in the azimuthal momentum equation.
We now wish to study the role of the radial drag term.
To keep the focus on the drag terms in the momentum equation, the parameters $\QT$, $\QR$,
and $\Dh$ will be fixed while $\taus$ will be varied.
For simplicity, we will set $\QT = \QR$ which corresponds to the case where the disk height
is estimated as $h_\rmp = c_\rmp/\Omega_0$.

To begin with let us set the mass diffusivity $\Dh = 0$
and plot the non-dimensional growth-rate $\mathrm{Re}(\gamma)$ (maximized with respect 
to wavenumber and the three roots of the cubic)
versus $\taus$; see Figure \ref{fig:qtqr}. 
The first case, $\QT = \QR = 0.2$ (Figure \ref{fig:qtqr}a), is unstable in the 
limit of zero explicit gas drag 
($\taus \to \infty$) in the classical SGW way.  
Observe that in this case, increasing overall gas drag 
(lowering $\taus)$ \textit{reduces} the growth-rate (solid line).
If one artificially retains only the azimuthal
gas drag (dashed line; radial drag turned off) then the instability 
is enhanced in accordance with the discussion of the preceding paragraphs.  
If only radial drag is 
retained (chain dotted; azimuthal drag turned off) we see that the instability
is further weakened compared to the case with both drag terms active. 
We thus conclude 
that azimuthal/radial drag promotes/hinders instability.  This is easy to 
appreciate: radial gas drag slows radial compressive and rarefactive particle motions.

The second case, $\QT = \QR = 2$ (Figure \ref{fig:qtqr}b), is neutrally stable without
gas drag ($\taus \to \infty$) and with $\Dh = 0$, i.e., neutrally stable in the SGW limit.  
It is only for this case that one may
say ``gas drag assists instability.'' 
Optimal enhancement occurs for $\taus = 1$.  If only 
azimuthal drag is present (dashed line, radial drag suppressed) then the more drag the better 
, i.e., the optimum disappears, again in agreement with the discussion in earlier paragraphs.
As $\taus$ decreases, radial drag diminishes the instability, while
azimuthal drag enhances it, leading to an optimum $\taus$.  
When only radial drag 
is present, the maximum growth-rate is zero and thus is not 
shown.  An inspection of all three roots of the cubic shows that this makes physical sense.  In the
absence of gas drag there is a zero root and two purely oscillating roots.  
In the presence
of radial drag alone the two oscillating roots are damped, while the zero root remains zero.

Next, let us introduce a little diffusivity, $\Dh = 1.0$.  The case $\QT = \QR = 0.2$
(SGW unstable) is shown in figure \ref{fig:qtqr}c.  Diffusivity does reduce growth-rates
(compare with figure \ref{fig:qtqr}a) as expected, however, above a certain $\taus$ ($\approx 2$ for 
this case) azimuthal drag is stabilizing.  
This is contrary to our physical explanation and it will be suggested below that a 
gravito-diffusive instability mechanism is at play. 
Inspection of the individual roots showed that this behavior always corresponded to
a pair of complex conjugate $\gamma$ roots with
non-zero oscillation frequency (overstability, $\mathrm{Im}(\gamma) \neq 0$).
This can also be described as a pair of growing waves propagating
radially inward and outward.
The bump or local maximum in the solid line near $\taus \approx 1$ shows that 
gas drag assisted behavior is present to the left of the plot.  This was not
the case for $\Dh = 0$.  
That is, mass diffusivity allows
the gas drag assisted instability to occur at lower $\QT$.
If we add mass diffusivity to the case ($\QT = \QR = 2$) that is gas drag assisted without
diffusivity, we observe (figure \ref{fig:qtqr}d) the same qualitative behavior to the left
of the plot as in $\Dh = 0$ case (apart from an expected decrease in peak growth-rate).
However, to the right of the plot, the gravito-diffusive
mode appears for $\taus$ above $\approx 200$, and closer inspection reveals that
for $\taus \gtrsim 400$ the diffusive case has a \textit{larger} growth-rate that the
non-diffusive case.  Notice that the gravito-diffusive mode survives in the gas-free 
limit $\taus \to\infty$ when there is
some mechanism of mass diffusivity still active to keep $\Dh$ finite.
An overstability with momentum diffusivity (viscosity) is known to exist in the context of planetary rings
(for a review see \cite{Schmidt_etal09}).  In the present case, we have encountered overstability with mass diffusivity.  We shall
not speculate on whether it is relevant in other astrophysical contexts.
Varying $\QT$ in the $\taus \to \infty$ limit (for the razor thin case, $\QR = 0$) shows (see figure \ref{fig:g_vs_QT}) that 
diffusivity allows the Toomre instability to occur at all $\QT$, albeit with
decreasing growth rate as $\QT$ increases.  It is for these reasons that we believe
this mode (which is overstable in nature) to be gravito-diffusive in nature and invite the reader to explain it
mechanistically. 

Since $\Dh$ decreases with increasing $\taus$ in the case of stirring by a gas,
the question arises whether the gravito-diffusive mode can be realized for particles
in a solar nebula.  To investigate this, a search was conducted in the range
$10^{-8}\le \alpha_\rmg \le 10^{-3}$, $0.1 \le a \le 100$ cm, and $0.1 \le R \le 100$ AU.
The most amplified mode never had $\mathrm{Im}(\gamma) \ne 0$. 

Figures \ref{fig:qtqr}e and \ref{fig:qtqr}f show that when
$\Dh = 100$ the distinction between high $\QT$ and low $\QT$ disappears.
Both now display gas drag assisted behavior except at sufficiently large
$\taus$ where the gravito-diffusive mode appears.

A different mechanistic interpretation of the instability has been
given by \cite{Goodman_and_Pindor00} as follows.  Consider a localized
clump of positive density perturbation in the particle layer in the
form of an axisymmetric ring.  At the inner edge of the ring,
particles will feel an extra outward gravitational force and, to
maintain radial equilibrium in absence of the gas, will need to travel
at less than Keplerian speed.  In the presence of gas the particles will therefore be 
energized by gas drag and
drift toward the center of the ring. At the outer edge of the ring,
particles will feel an extra inward gravitational force and will need
to travel at faster than Keplerian speed to maintain radial
equilibrium.  If gas drag is now turned on, they will be slowed down
and drift inward.

Compared to our explanation, that of \cite{Goodman_and_Pindor00}
assumes equilibrium between gravitational forces and centripetal
acceleration, i.e., it assumes zero restoring force due to the angular
momentum gradient.  This holds in the limit (very often true and discussed previously) of
perturbation time scale much smaller than the particle stopping time.

Finally, it is worth emphasizing that, unlike particle concentration
in pressure highs
\citep{Barge_and_Sommeria95,Haghighipour_and_Boss03a} which occurs
solely due to gas drag, the present instability requires both
self-gravity and gas drag.

\section{Results}

Table \ref{tab:models} summarizes the five models considered in paper II and
repeated here with the mass diffusivity term added.
The reference model has the following parameter values: particle
size, $a = 1$ mm, enhancement/depletion factors (relative to the minimum mass
solar nebula) of gas, $f_\rmg = 1$, and of particles, $f_\rmp = 1$.  Figure
\ref{fig:tgrow_reference} is analogous to Figure 4 (top) in paper II and shows 
e-fold growth times for the reference case and various weak to moderate values of
$\alpha_\rmg$. (Here and henceforth, growth times are plotted for the fastest
mode among all three roots and all wavenumbers.) The lower set of curves (of lighter weight) are for
zero mass diffusivity.  The upper
curves (heavier weight) are for the case with diffusivity.  One concludes that including mass diffusivity
increases growth times by about two orders of magnitude in the asteroid belt
region at 3 AU.  It appears that in this region turbulence levels must be such
that $\alpha_\rmg < 10^{-7}$ to get growth times smaller than $10^6$ years, a
typical disk lifetime, and the situation becomes worse in the terrestrial
planet region. 
However, the instability could perhaps play a role in the outermost
regions of the disk for weak turbulence levels.

For the same turbulence levels as in Figure \ref{fig:tgrow_reference}, we now 
consider the effect of particle
size $a$ at $R = 3$ AU; see Figure \ref{fig:tgrow_versus_size}.
Growth time (heavy lines) decreases for larger particles.
Also shown is the characteristic drift time of the instability wave which also decreases with
particle size but more slowly than the growth time giving rise to
a possibility that $t_\mathrm{grow} < t_\mathrm{drift}$.  One observes that
such a cross over occurs (for particle radii in the range of the plot) when the turbulence levels is sufficiently
weak.  Furthermore, the cross-over occurs for smaller particles as the turbulence diminishes.

Paper II also considers the following four alternative models
each of which makes the instability stronger: (i) an increase
in the surface density of particles by a factor of $f_\rmp = 4$, (ii) a depletion in
gas by a factor of $f_\rmg = 0.1$,  (iii) an increase in the particle radius to $a =
1$ cm, and, finally, (iv) a model designated ``all'' that incorporates all of these
changes.

The ``all'' model, in which solids are enhanced relative to gas by a factor of 40 compared
to the minimum mass solar nebula, produces the fastest growth times of all the models: these
are plotted in figure \ref{fig:tgrow_all} for larger turbulence intensities $\alpha_\rmg$ than in the previous
figure up to the (global) value $\alpha_\rmg \sim 0.01$ required to account for observed disk
accretion rates.  The
increase in growth time at 3 AU from the non-diffusive to the diffusive
treatment is now just less than an order of magnitude.  Thus a solids-enhanced nebula is much less
affected by turbulent diffusivity. This is not surprising because 
in the limit of vanishing gas density, none of the bothersome gas-related 
obstacles to GI will occur.  For the standard turbulence intensity of $\alpha_\rmg = 0.01$
the instability is not strong enough (e-fold time barely less than a million years).
It should be noted that unlike the reference model, the ``all'' model 
produces a peak in growth rate
at about 7 AU.

For the same turbulence levels as in Figure \ref{fig:tgrow_all} we now consider
the effect of particle size.
Figure \ref{fig:tgrow_versus_size_enriched} shows growth times (heavy lines) versus 
particle size for the particle enriched and gas depleted case ($f_\rmg = 0.1$, 
$f_\rmp = 4$).  Again, these should be compared with the thin lines which show the
corresponding drift time.  One concludes that there exists a range of particles
sizes for which $t_\mathrm{grow} < t_\mathrm{drift}$ only if the turbulence is weaker
than $\alpha_\rmg \lesssim 10^{-3}$. 

%Figure \ref{fig:tgrow_all_new} shows growth times for values
%of turbulence intensity $\alpha_\rmg = 10^{-5}$ to $10^{-3}$.
%Interestingly, the growth-rate peaks at $R \approx 8$ AU.

Paper II defines a ``viable'' instability as the satisfaction of three
conditions.  (1) e-folding times reasonably smaller than the disk lifetime,
in particular $t_\mathrm{grow} < 10^5$ yrs.  (2) Growth times less than the
drift time: $t_\mathrm{grow} < t_\mathrm{drift}$ where $t_\mathrm{drift}$ is given in
equation (16) of paper II.  (3) Wavelength of the fastest mode reasonably
shorter than the radius: $\lambda_\mathrm{f} < R/2$.  Figure \ref{fig:alpha_max} shows
the maximum value of turbulence intensity $\alpha_\rmg$ that is still able to
produce a viable gravitational instability of solids. The condition that is
violated when $\alpha_\rmg$ exceeds the maximum value is depicted
using different line types: solid, dashed, and dotted for the three conditions,
respectively.  One can see that under the favorable conditions represented by
the ``all'' model, the instability becomes non-viable in the terrestrial planet
region when the turbulence level exceeds $\alpha_\rmg = 10^{-4}$, and
everywhere when $\alpha_\rmg > 10^{-3}$.

Finally, Figure \ref{fig:gamma_approx} shows the relative error in the maximum
growth rate given by the approximate formula \eqp{approx} valid for small
growth rates compared to stopping times. The value plotted is the relative
error in the maximum growth rate with respect to wavenumber. For the two cases
shown, the approximate formula gives answers correct to within 10\% if $\gamma_\mathrm{approx}
\tau_\mathrm{s} \lesssim 10^{-3}$.
\section{Conclusion}

Under conditions of the minimum mass solar nebula at the distance of the
terrestrial planets and the asteroid belt, the gas-drag mediated
instability remains viable for chondrule sized particles only for very weak turbulence levels
$\alpha_\rmg < 10^{-8}$.  Even under the optimistic conditions of
enhancement of the ratio of particle column density to gas column density by a
factor of 40, and 1 cm particles (both of which promote the instability) the
instability remains viable only if $\alpha_\rmg < 10^{-4}$. Things look a
little more optimistic further out.  At 10 AU, under standard conditions the
instability becomes too slow compared to nebula lifetime when $\alpha_\rmg >
10^{-7}$ while under the most optimistic conditions, viability of the
instability is destroyed by exceedingly large wavelength for $\alpha_\rmg > 2 \times
10^{-4}$.  Clearly, research should be undertaken to pin down turbulence levels at
the disk mid-plane.

\newpage
\centerline{\bf Acknowledgement}

We would like to thank the referee for useful suggestions and pointing out
several typographical errors and one logical error in the original manuscript.
We also thank Dr. A. Youdin for suggestions that improved the paper.

%%%%%%%%%%%%%%%%%%%%%%%%%%%%%%%%%%%%%%%%%%%%%%
\bibliography{paper}
\bibliographystyle{plainnat}

\clearpage

\begin{table}\begin{center}
\begin{tabular}{|c|c|c|c|}
\hline
Name            & particle size, $a$ & gas ratio, $f_\rmg$ & particle ratio, $f_\rmp$ \\ \hline
reference       & 1 mm               & $1.0$               & $1.0$                    \\
$f_\rmg = 0.1$  & 1 mm               & $0.1$               & $1.0$                    \\              
$f_\rmp = 4$    & 1 mm               & $1.0$               & $4.0$                    \\              
$a = 1$ cm      & 1 cm               & $1.0$               & $1.0$                    \\
all             & 1 cm               & $0.1$               & $4.0$                    \\
\hline
\end{tabular}
\caption{Models}
\label{tab:models}
\end{center}\end{table}

%%%%%%%%%%%%%%%%%%%%%%%%%%%%%%%%%%%%%%%%%%%%%%%%%%%%%%%
\begin{figure} % 1
\vskip 0.50truecm
\begin{center}\includegraphics[width=4.0truein]{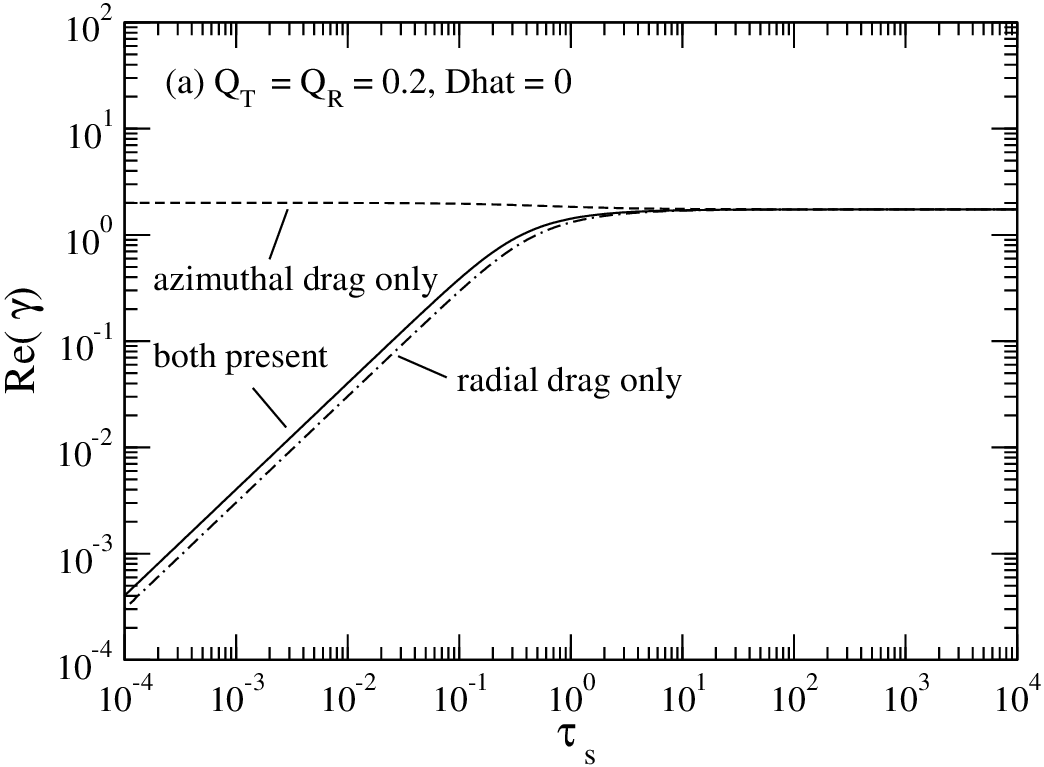}\end{center}
\vskip 0.50truecm
\begin{center}\includegraphics[width=4.0truein]{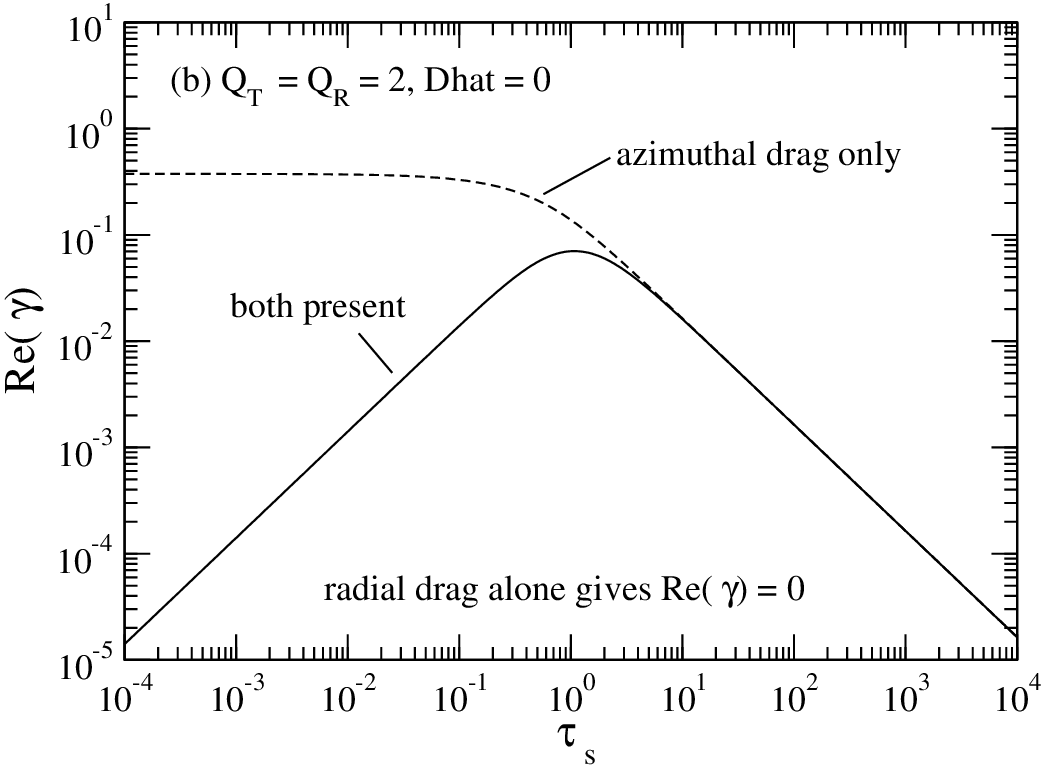}\end{center}
\end{figure}
\newpage

\begin{figure}
\vskip 0.50truecm
\begin{center}\includegraphics[width=4.0truein]{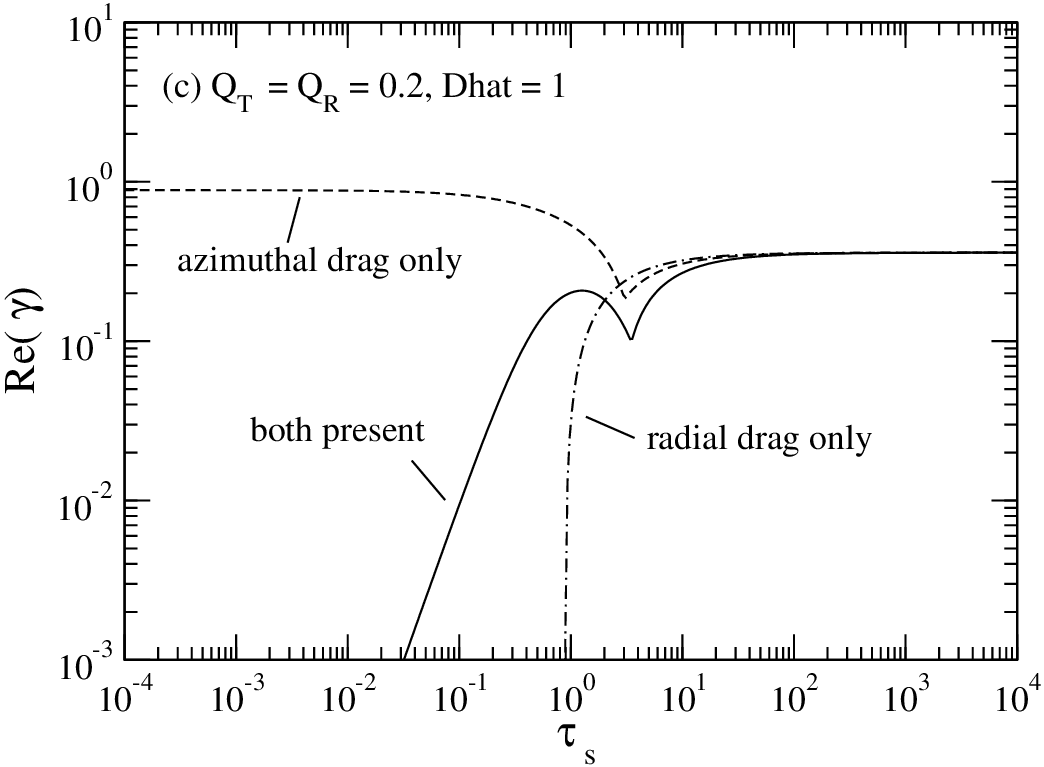}\end{center}
\vskip 0.50truecm
\begin{center}\includegraphics[width=4.0truein]{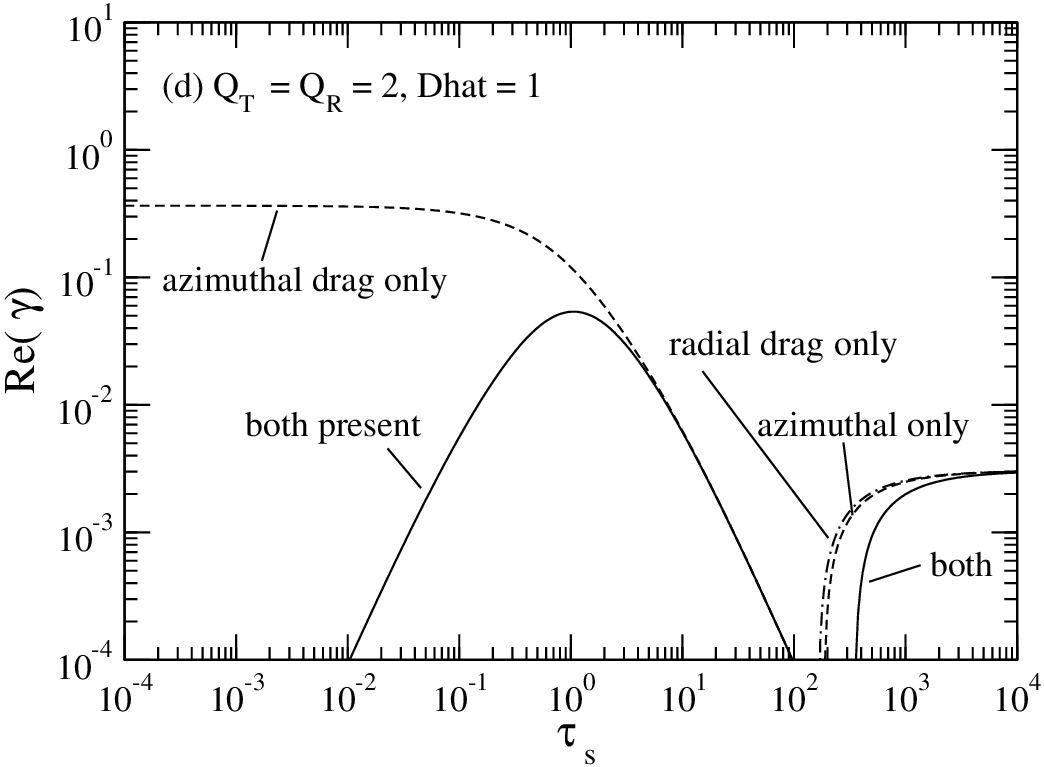}\end{center}
\end{figure}
\newpage

\begin{figure}
\vskip 0.50truecm
\begin{center}\includegraphics[width=4.0truein]{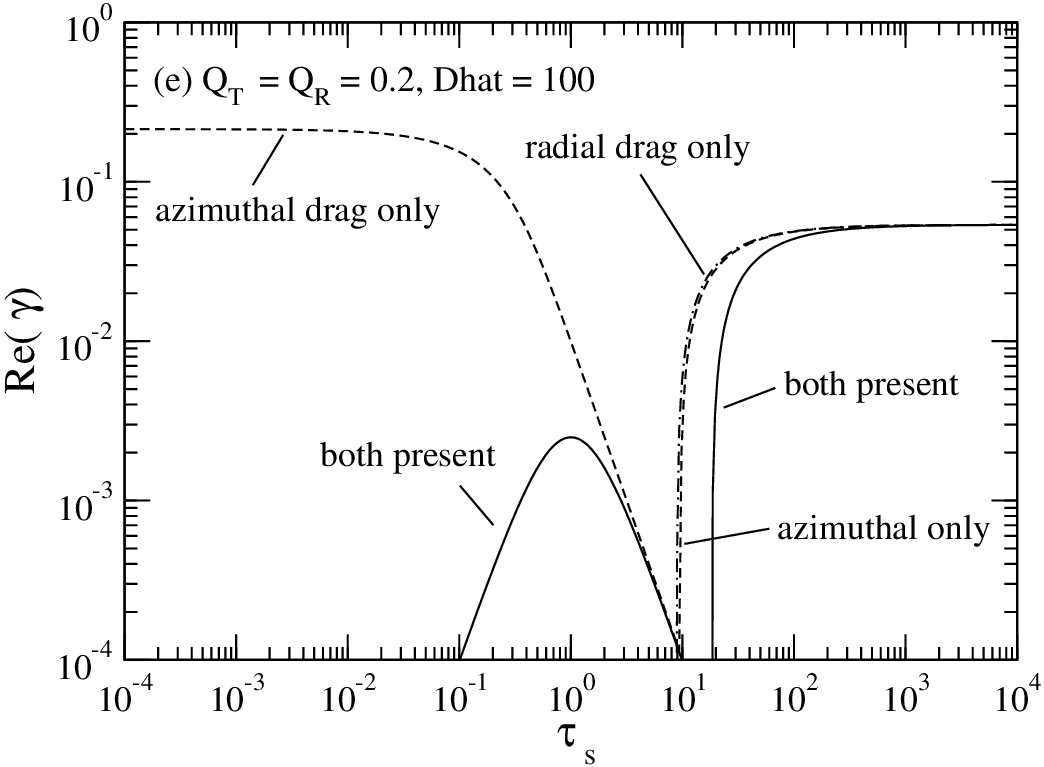}\end{center}
\vskip 0.50truecm
\begin{center}\includegraphics[width=4.0truein]{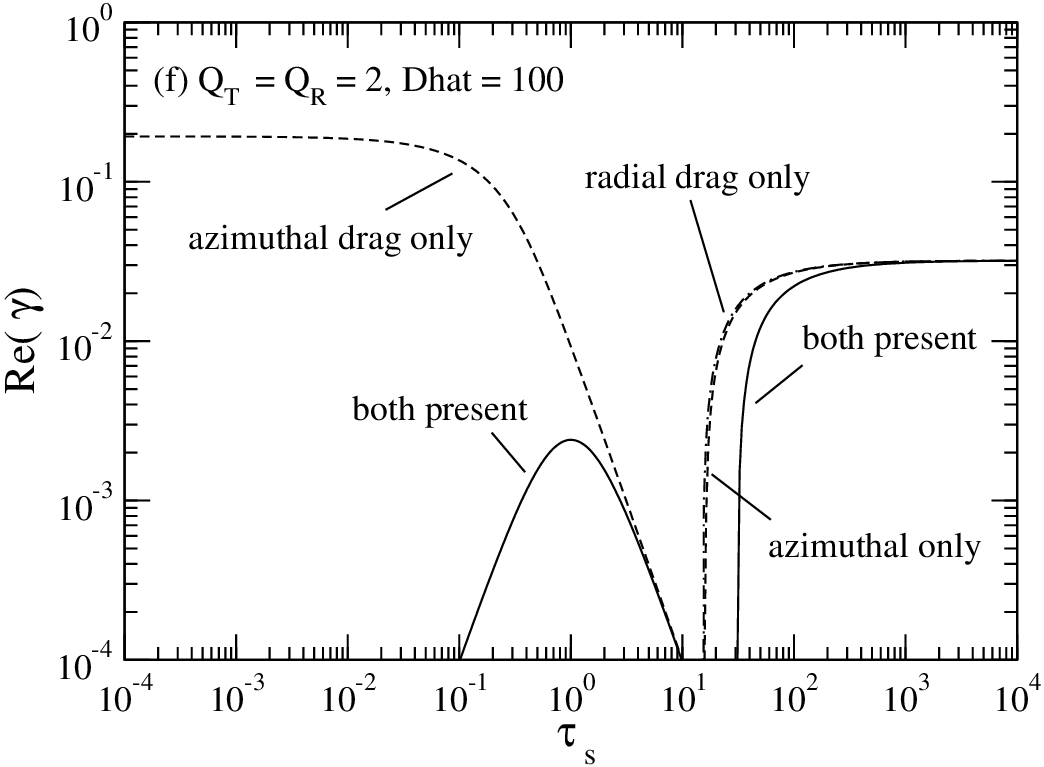}\end{center}

\caption{Study of the effect of the explicit drag terms in the momentum equation.
Growth rate $\mathrm{Re}(\gamma)$ versus $\taus$ at fixed $\QT, \QT$, and $\Dh$.
\solid, both azimuthal and radial drag terms active;
\dashed, only azimuthal drag active;
\chndot, only radial drag active.}
\label{fig:qtqr}
\end{figure}
%%%%%%%%%%%%%%%%%%%%%%%%%
\begin{figure}\begin{center} % 2
\includegraphics[width=5.0truein]{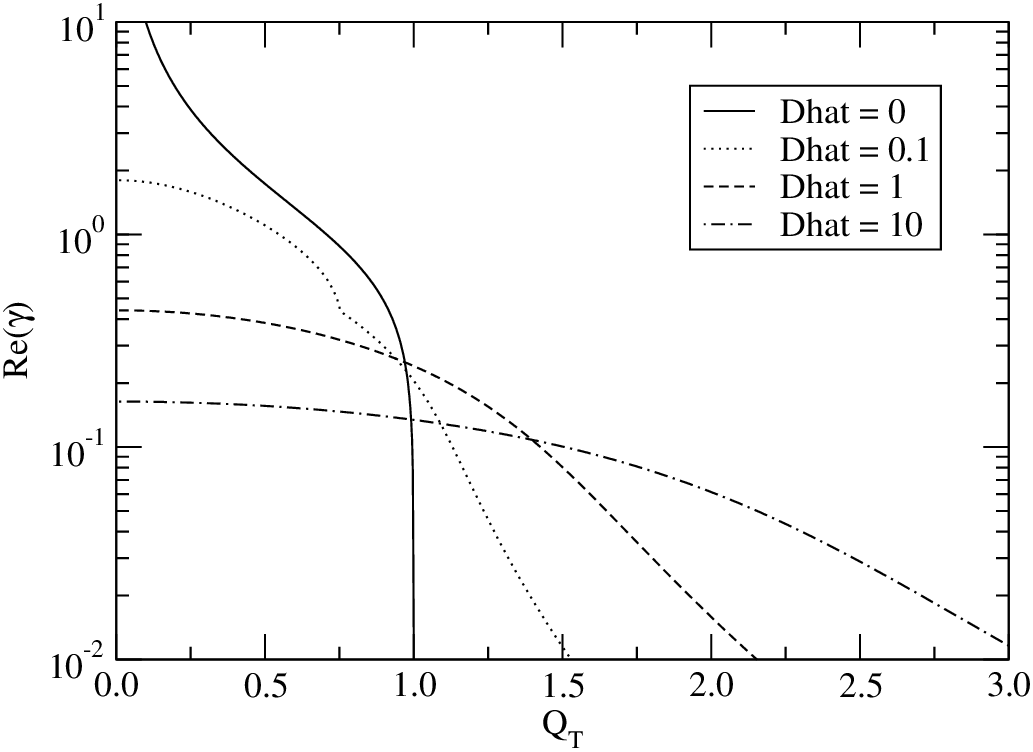}
\caption{Maximum growth rate $\mathrm{Re}(\gamma)$ versus the Toomre parameter in the drag free limit with finite
mass diffusivity $\Dh$.  The razor thin disk, $\QR = 0$, is being considered.
\solid,  $\Dh = 0$;
\dotted, $\Dh = 0.1$;
\dashed, $\Dh = 1$;
\chndot, $\Dh = 10$.}
\label{fig:g_vs_QT}
\end{center}\end{figure}
%%%%%%%%%%%%%%%%%%%%%%%%%%
\begin{figure}\begin{center} % 3
\includegraphics[width=5.0truein]{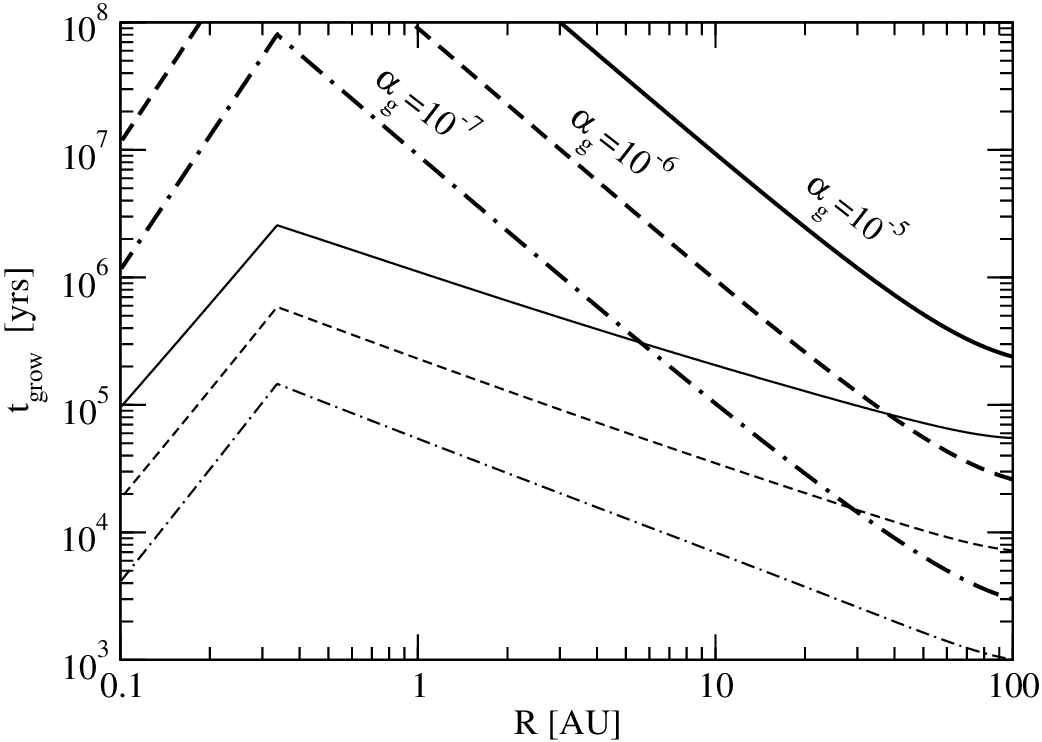}
\caption{Growth times (mimimized with respect to wavelength) for the reference model.  
Two curves are shown for each line
type.  The lower set of curves (light type) is for zero mass
diffusivity while the upper set (heavy type) is for non-zero
diffusivity.  Each line type corresponds to a different value of $\alpha_\rmg$ as
follows:
\solid,  $\alpha_\rmg = 10^{-5}$;
\dashed, $\alpha_\rmg = 10^{-6}$;
\chndot, $\alpha_\rmg = 10^{-7}$.
Note: Growth times above $10^8$ yr are not displayed.
}
\label{fig:tgrow_reference}
\end{center}\end{figure}
%%%%%%%%%%%%%%%%%%%%%%
\begin{figure}\begin{center} % 4
\includegraphics[width=5.0truein]{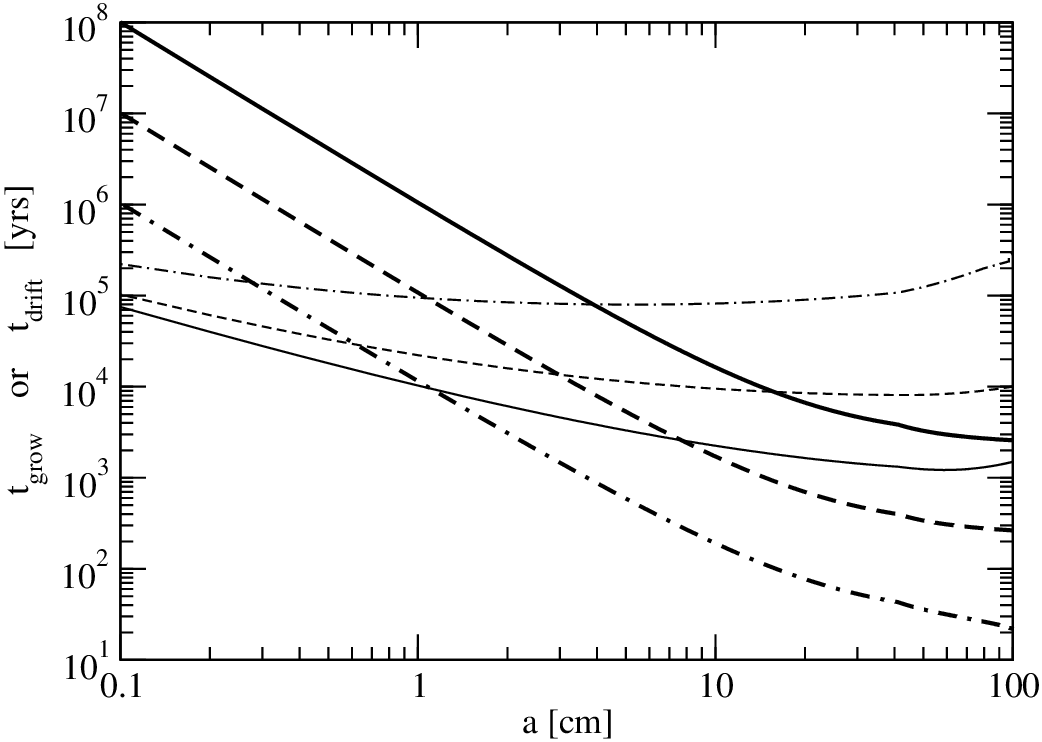}
\caption{Effect of particle radius $a$ on growth time (heavy lines) which
should be compared with the corresponding characteristic drift time (thinner lines)
of the instability wave.  Mass diffusivity has been included.
Minimum mass nebula ($f_\rmg = 1$, $f_\rmp = 1$) at 3 AU.  The same turbulence 
strengths and line types
as Figure \ref{fig:tgrow_reference} are used, namely:
\solid,  $\alpha_\rmg = 10^{-5}$;
\dashed, $\alpha_\rmg = 10^{-6}$;
\chndot, $\alpha_\rmg = 10^{-7}$.
}
\label{fig:tgrow_versus_size}
\end{center}\end{figure}
%%%%%%%%%%%%%%%%%%%%%%
%%%%%%%%%%%%%%%%%%%%%
\begin{figure}\begin{center}% Fig. 5
\includegraphics[width=5.0truein]{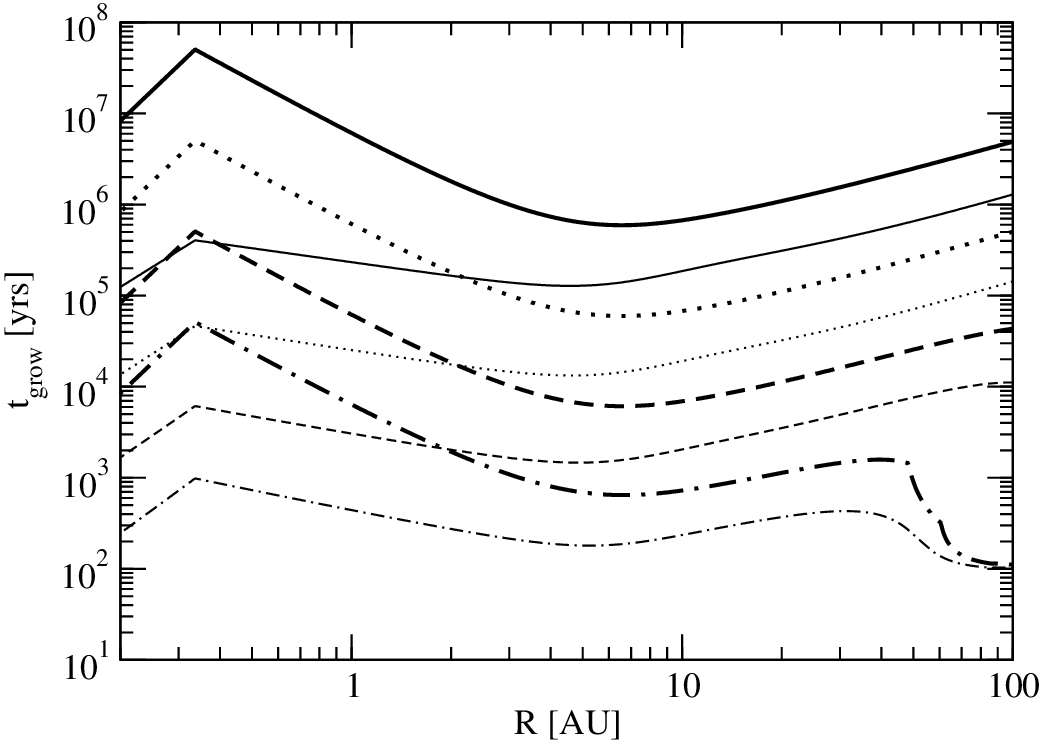}
\caption{Growth times for model ``all'' and different turbulence strengths $\alpha_\rmg$ as follows:
\solid, $\alpha_\rmg = 0.01$;
\dotted,  $\alpha_\rmg = 10^{-3}$;
\dashed, $\alpha_\rmg = 10^{-4}$;
\chndot, $\alpha_\rmg = 10^{-5}$.
The curves with heavy lines are for non-zero mass diffusivity while
the thin lines are for zero mass diffusivity.  The abscissa starts at $R=0.2$ to ensure that
the particle Reynolds number $\le 1$.}
\label{fig:tgrow_all}
\end{center}\end{figure}
%%%%%%%%%%%%%%%%%%%%%%
\begin{figure}\begin{center} % Fig. 6
\includegraphics[width=5.0truein]{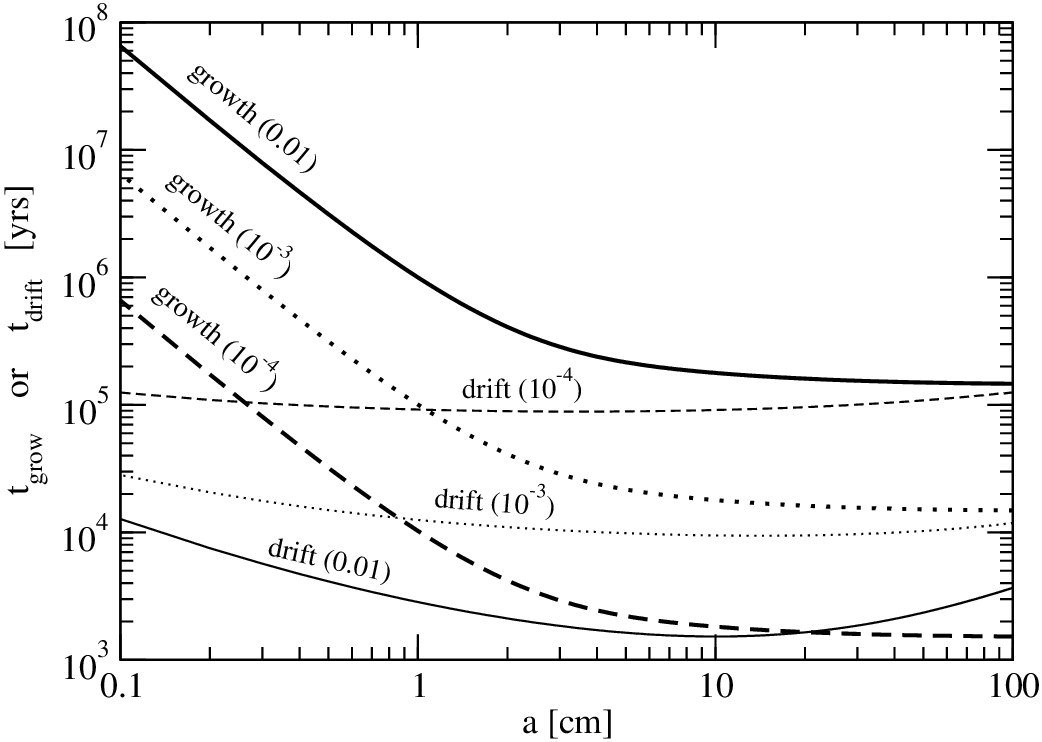}
\caption{Effect of particle size $a$ on growth time (heavy lines)  which
should be compared with the corresponding drift time (thinner lines).  
Particle enriched and gas depleted nebula ($f_\rmg = 0.1$, $f_\rmp =
4$) at 3 AU.  The same turbulence strengths and line types
as Figure \ref{fig:tgrow_all} are used, namely:
\solid,  $\alpha_\rmg = 0.01$;
\dotted, $\alpha_\rmg = 10^{-3}$;
\dashed, $\alpha_\rmg = 10^{-4}$.}
\label{fig:tgrow_versus_size_enriched}
\end{center}\end{figure}
%%%%%%%%%%%%%%%%%%%%
\begin{figure} % Fig. 7
\begin{center}\includegraphics[width=4.5truein]{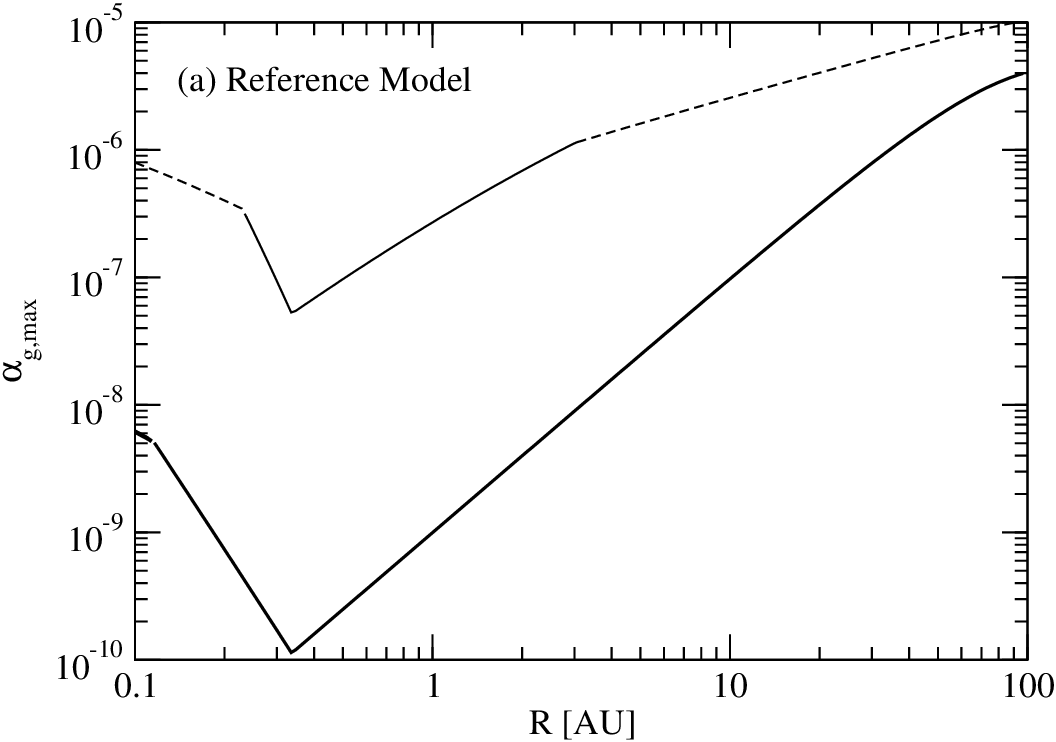}\end{center}
\vskip 1.0truecm
\begin{center}\includegraphics[width=4.5truein]{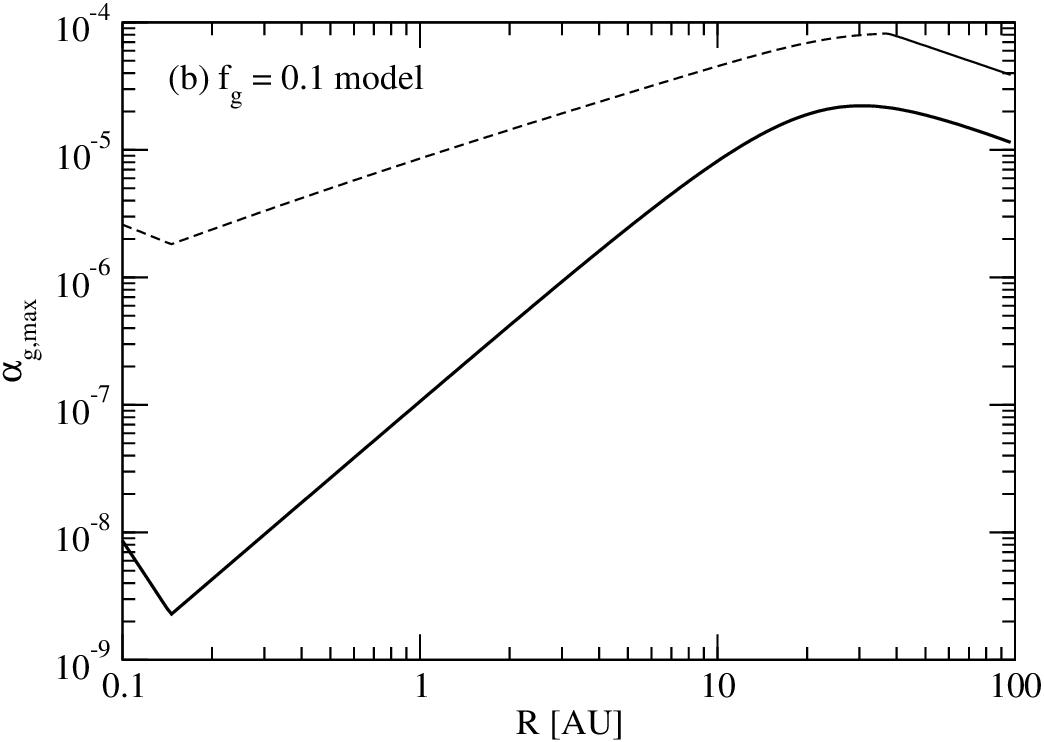}\end{center}
\end{figure}
\newpage

\begin{figure}
\begin{center}\includegraphics[width=4.5truein]{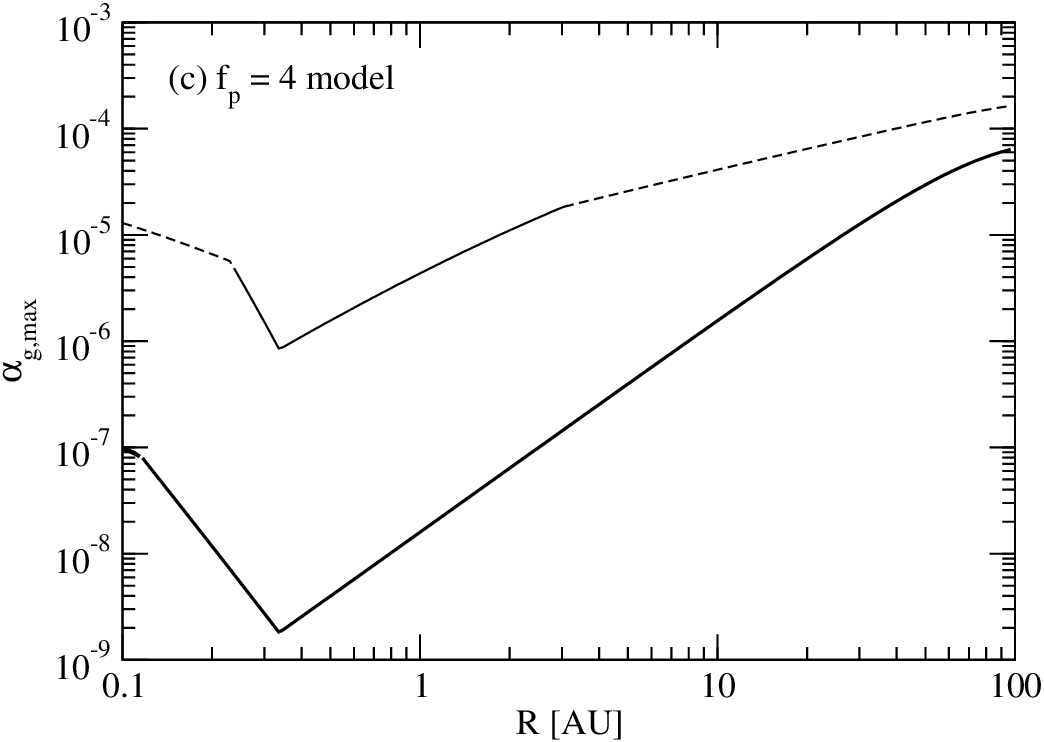}\end{center}
\vskip 1.0truecm
\begin{center}\includegraphics[width=4.5truein]{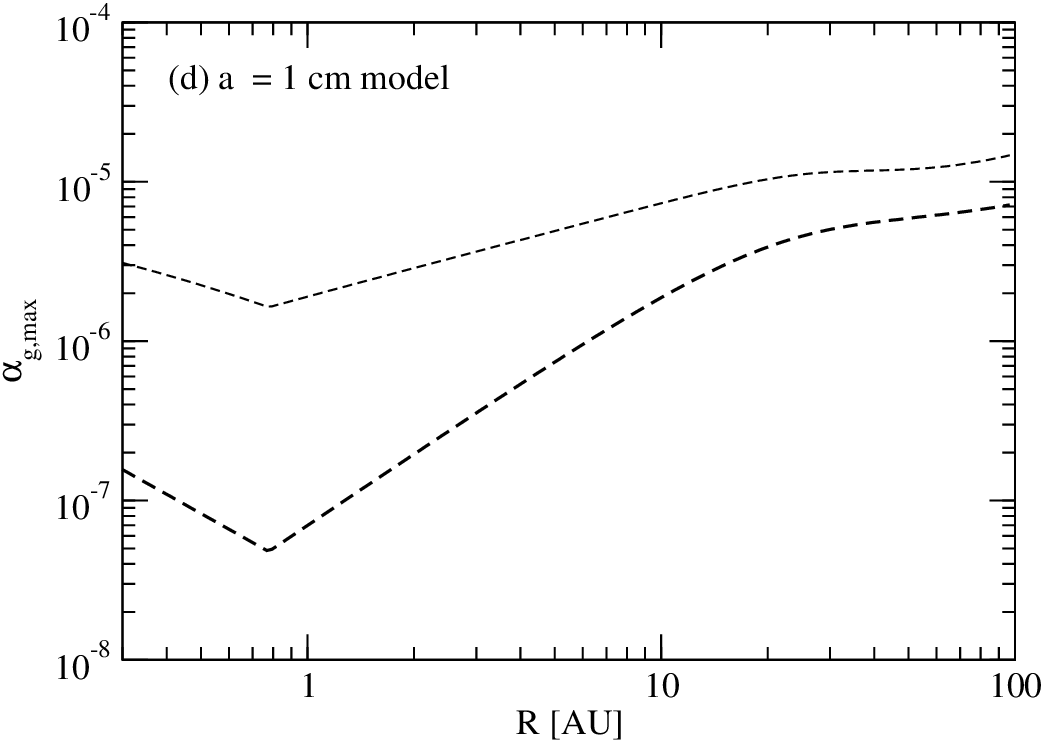}\end{center}
\end{figure}

\begin{figure}
\begin{center}\includegraphics[width=4.5truein]{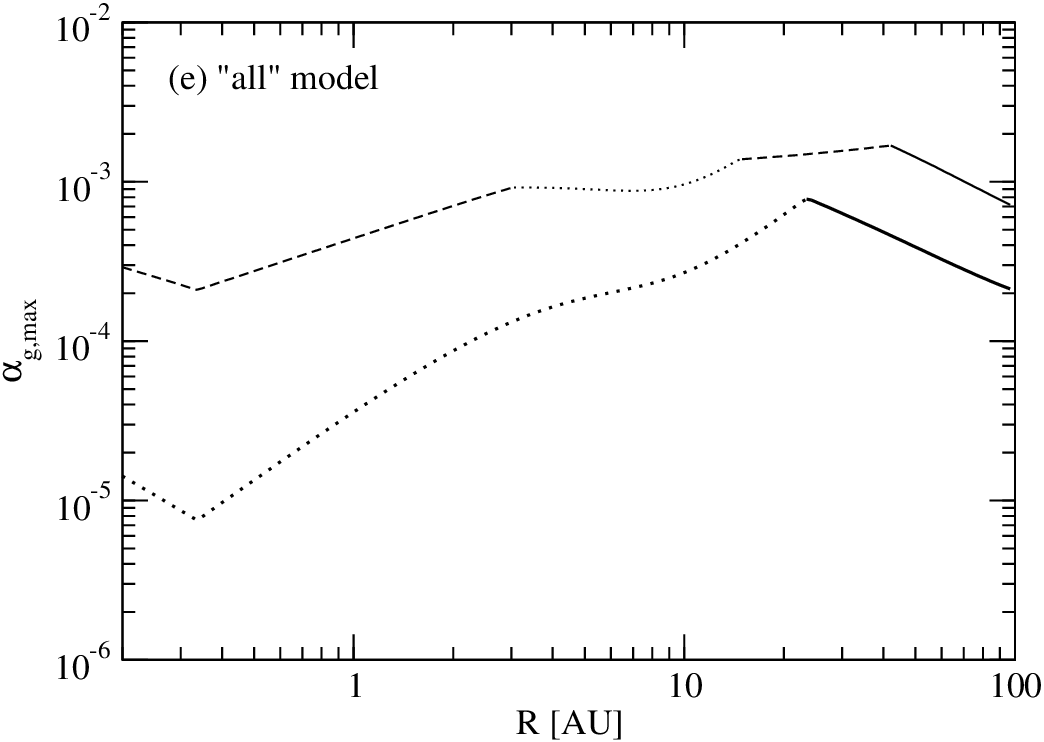}\end{center}
\caption{The maximum turbulence level $\alpha_\rmg$ the instability can tolerate.  
There are two curves for each model.  The upper curve (light weight) is for
zero mass diffusivity while the lower one (heavy weight) is for non-zero diffusivity.
The line-type denotes the condition that sets the maximum $\alpha_\rmg$:
\solid: nebula lifetime;
\dashed: radial drift;
\dotted: wavelength.
The abscissa starts at $R = 0.3$ AU in (d) and $R = 0.2$ AU in (e) to ensure a particle Reynolds number $\ge$ 1.}
\label{fig:alpha_max}.  
\end{figure}
%%%%%%%%%%%%%%%%%%%%%%%%%%%%%%
%%%%%%%%%%%%%%%%%%%%
\begin{figure} % Fig. 8
\begin{center}
\includegraphics[width=4.5truein]{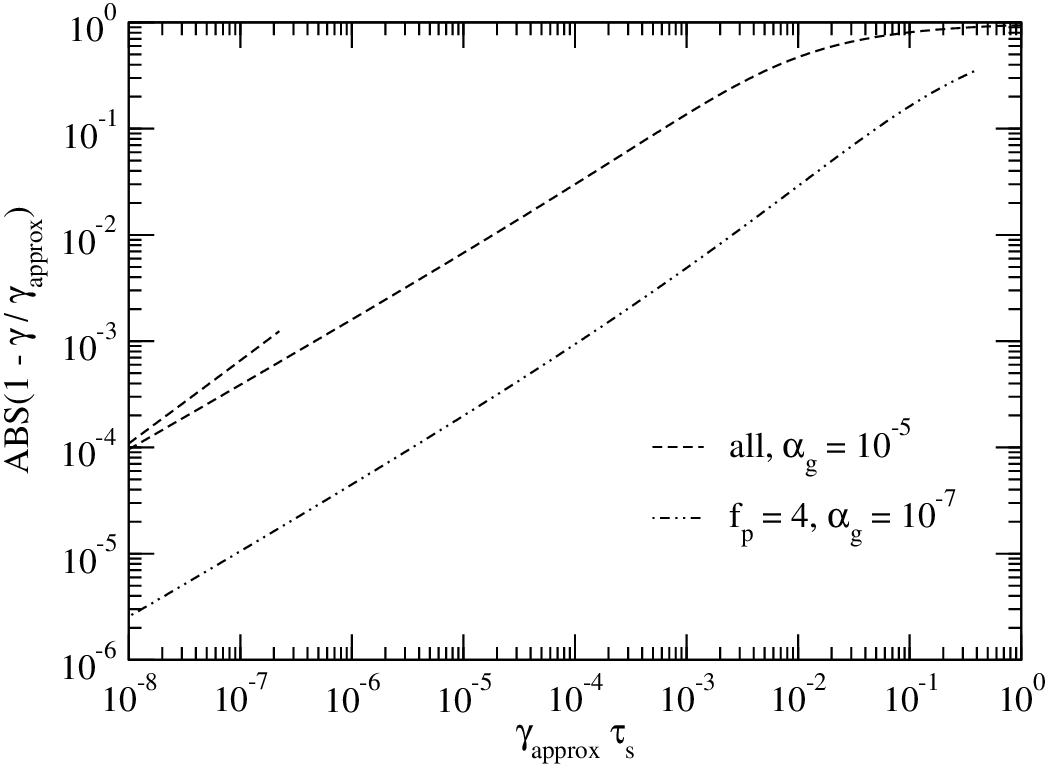}
\caption{Locus of the relative error in the growth-rate (maximized over wavelength) predicted by the approximate formula
\eqp{approx} as compared with the actual value (separately maximized over wavelength).  Each locus was obtained
by varying the radius $R$ and plotting the relative error versus $\gamma_\mathrm{approx}\taus$.
$D \neq 0$ for both cases.  
\chnddot: ``all'' model and $\alpha_\rmg = 10^{-5}$;
\dashed, $f_\rmp = 4$ model and $\alpha_\rmg = 10^{-7}$.}
\label{fig:gamma_approx}
\end{center}
\end{figure}
%%%%%%%%%%%%%%%%%%%%%%%%%%%%%%%%%%%%%%%%%%%%%%%%%%%%%%
%%%%%%%%%%%%%%%%%%%%%%%%%%%%%%%%%%%%%
%\begin{figure} % Fig. 10
%\begin{center}
%\includegraphics{slice.eps}[width=5truein]
%\caption{Growth times along a horizontal cut at $a=20$ cm through the diamonds ($\alpha_\rmg = 10^{-6.5}$) of the previous figure.
%Drift times are well above growth times.}
%\label{fig:slice}
%\end{center}
%\end{figure}
%%%%%%%%%%%%%%%%%%%%%%%%%%%%%%%%%%%
%\begin{figure} % Fig. 10
%\begin{center}
%\includegraphics{slice-4.eps}[width=5truein]
%\caption{Growth time for $a = 20$ cm when $\alpha_\rmg = 10^{-4}$.}
%\label{fig:slicemf}
%\end{center}
%\end{figure}
%%%%%%%%%%%%%%%%%%%%%%%%%%%%%%%%%%%%
%\begin{figure}\begin{center}
%\includegraphics{michikoshi_2011.eps}[width=5truein]
%\caption{Effect of turbulence intensity $\alpha_\rmg$ for the
%  conditions of \cite{Michikoshi_etal10}, namely, a minimum mass
%  nebula at 1 AU.
%heavy \solid line : growth time for 20 cm particles;
%thin \solid line : the corresponding drift time;
%heavy \dotted line : growth time for 1 cm particles;
%thin \dotted line : the corresponding drift time.}
%\label{fig:michikoshi}
%\end{center}\end{figure}
\end{document}